\documentclass[aps,amsmath,amssymb,showpacs]{revtex4}
\usepackage{color}
\usepackage{graphicx}

\begin{document}

\title{A Kolmogorov proof of the Clauser, Horne, Shimony and Holt inequalities }

\author{M. Revzen}
\affiliation {Department of Physics, Technion - Israel Institute of Technology,
Haifa 32000, Israel}

\date{\today}

\begin{abstract}

Boolean logic is used to prove the CHSH inequalities. The proof elucidates the
connection between Einstein elements of reality and quantum non locality.
The violation of the CHSH inequality by quantum theory is discussed and the two stage view of quantum measurement relevance to incompatible observables is outlined.

\end{abstract}

\pacs{XXX}

\maketitle

\section{Introduction}

A convenient specification of Einstein, Podolsky, Rosen and Bohm experimental set up that we adopt is:
An experiment is made of numerous runs. In each run two photons (1,2) in the state $|\psi(1,2)>= \frac{1}{\sqrt{2}}(|+_1+_2>+|-_1-_2>)$ are involved. (Here +(-) means positive (negative) polarization along a common z axis.) The photons propagates to two separated ports
($\alpha, \beta$) therein their polarizations are measured  by one of two polarizers:  A or A' at
$\alpha$, and B or B' at $\beta$. The experiments are assumed flawless: In each run both photons reach the counters
at $\alpha$ and $\beta$ and record a reading at both: +1 if the photon pass the polarizer and -1 if it doesn't. \\

The issue under study is: can the experimental results of the EPRB set up be accounted for in terms of canonical (Kolmogorov's) probability theory
based on hidden variable, $\lambda$?\\

Two point correlation of the outcomes with the polarizers set at A and B, $<AB>$, is defined by:\\

\begin{equation}\label{th}
<AB>=\frac{1}{N}\sum_{i=1}^{N}a_ib_i,
\end{equation}

Here i enumerates the run and a,b the outcome at the ports $\alpha$, $\beta$ respectively.

An account by probability theory based on hidden variable implies that two point correlations may be expressed by

\begin{equation}\label{hv}
<AB>=\int d\lambda\rho(\lambda)A(\lambda)B(\lambda),
\end{equation}

with a distribution, $\rho(\lambda),$ common to all correlations, $\lambda$ is the "hidden variable".

We note that within such an account the correlations $<AA'>$ and $<BB'>$ are defined though are not measured (perhaps even not measureable
e.g. due to technical difficulties). Thus, e.g.,  \\

\begin{equation}\label{simul}
 <AA'>=\int d\lambda\rho(\lambda)A(\lambda)A'(\lambda).
\end{equation}

The canonical (Kolmogorov's) probability theory involves definition of a sample space wherein each of the points accounts for possible
experimental outcomes. The total sample space, $\Omega$, is a set of points that cover all possible experimental outcomes. The  measure assigned to these points is their probability. Observable such as e.g. $A(\lambda)$, are termed events. Our analysis does not require the specification of this space - suffice it to know that such exists for a canonical probability theory. This allows us
to analyse interrelations among events via the algebra of sets aided by intuitively appealing  Venn diagrams.\\

In the next section we obtain, assuming the existence of a probability theory and using simple Boolean logic \cite{pitowsky,hess,khren} implied consistency  relations among
the two points correlations it allows. These interrelations will then be shown to be the so called Bell's inequalities.\\

\section{Interrelations among Correlations in Probability Theory}

 In classical Kolmogorov theory interrelations among probabilities may be analysed
in terms of Venn diagrams pertaining to the corresponding events \cite{zvi},\cite{math}. Thus, e.g., we may consider the "area" in a Venn diagram "occupied"  by the event $(AB)_{=}$. This event relates to the probability of observing  equal readings, A=B, i.e. polarizer A (at $\alpha$) and B (at $\beta$)  both read +1 or both -1. In our notation (using A and B as an example) this means: $(AB)_{=}$ is the set of all the sample space points with A=B. Its measure is the  probability of this event, $P((AB)_{=}))$. The complementary event $(\widetilde{AB})_{=}$, containing the sample points with unequal values for A and B, is designated by $(AB)_{\times}$. Its probability is  denoted by $P((AB)_{\times})$.\\

Thus,\\

\begin{equation}\label{tilde}
(\widetilde{AB})_{=}=(AB)_{\times}\Rightarrow P((\widetilde{AB})_{=})\equiv  P((AB)_{\times}),
\end{equation}

and
\begin{equation}\label{P}
P\big[\big((AB)_{=}\cup (AB')_{=}\big)\big]+P\big[\widetilde{\big(AB)_{=}\cup (AB')_{=}\big)}\big]=1.
\end{equation}

Similar relations hold for the other two points correlations of interest, e.g., $(AB')_{=}, (A'B)_{=}, (A'B')_{=},$  etc.

Boolean logic, i.e. set algebra, dictates \cite{pitowsky},\cite{hess2},\cite{khren},\cite{math},

\begin{equation}\label{tilde2}
\widetilde{\big((AB)_{=}\cup (AB')_{=}\big)}=(\widetilde{AB})_{=}\cap(\widetilde{AB'})_{=}=(AB)_{\times}\cap(AB')_{\times}.
\end{equation}

Where we used Eq.(\ref{tilde}) in the last step.

We have trivially that $(AB)_{\times}\cap(AB')_{\times}\Rightarrow (BB')_{=}$ i.e. :\\

\begin{equation}
\big((AB)_{\times}\cap(AB')_{\times}\big)\subseteq(BB')_{=},
\end{equation}

implying \cite{it},

\begin{equation}\label{b1}
P((AB)_{=})+P((AB')_{=})+P((BB')_{=})\;\ge\; 1.
\end{equation}

Eq.(\ref{b1}) is a consistency requirement stemming from pairing the probabilities  of $(AB)_{=}$ with $(AB')_{=}$.\\

\begin{figure}[h]
  \def\svgwidth{0.40\columnwidth}%
  \resizebox{.4\columnwidth}{!}{\input{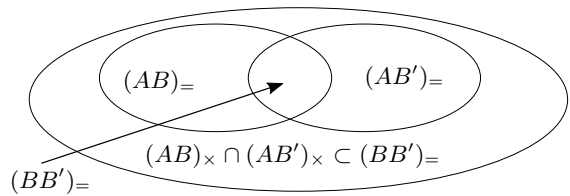}}%
  \caption{Probability Consistency Interrelation: $P((AB)_{=})+P((AB')_{=})+P(BB')_{=})\;\ge\;1$}
\end{figure}

Quite generally this approach allows derivation of consistency relations among  probabilities. These are equivalent to
Bell's inequalities which are formulated in terms of correlations. We now proceed to reformulate our consistency relations
in term of correlations.\\

\begin{eqnarray}\label{c1}
P((AB)_{=})+P((AB)_{\times})&=&1, \nonumber \\
P((AB)_{=})-P((AB)_{\times})&=& <AB>, \nonumber \\
\Rightarrow &&\nonumber \\
2P((AB)_{=})\;=1&+&<AB>, \nonumber \\
2P((AB)_{\times})\;=1&-&<AB>.
\end{eqnarray}

Utilizing Eq.(\ref{c1}), Eq.(\ref{b1}) may be written in terms of two point correlations as

\begin{equation}\label{cc1}
\langle AB\rangle +\langle AB'\rangle+\langle BB'\rangle\;\;\ge\; -1
\end{equation}

Going through similar reasoning with the pair $(AB)_{=}\cup(AB')_{\times}$  gives,

\begin{equation}\label{cc2}
\langle AB\rangle-\langle AB'\rangle-\langle BB'\rangle\;\;\ge\; -1
\end{equation}

The inequality for $(A'B)_{=}$ and $(A'B')_{=}$, is of course identical to Eq(\ref{cc1}) with A' replacing A:

\begin{equation}\label{cc3}
\langle A'B\rangle+\langle A'B'\rangle+\langle BB'\rangle\;\;\ge\; -1.
\end{equation}

Combining the inequality, Eq.(\ref{cc2}), with that of Eq.(\ref{cc3}) yields a Bell's inequality:

\begin{equation}\label{B1}
\langle AB\rangle-\langle AB'\rangle+\langle A'B\rangle+\langle A'B'\rangle\;\ge\; -2.
\end{equation}

There are 4 possible pairings:\;$ 1 \;(AB),(AB');\;2.\;(A'B),(A'B');\;3. \;(AB),(A'B);\;4.\; (AB'),(A'B')$.\\

There are 4 combinations for each, e.g.:\;1. \;$(AB)_{=},(AB')_{=};\;2.\;(AB)_{=},(AB')_{\times};\;3.\;(AB)_{\times},(AB')_{=};\;4.\;(AB)_{\times},(AB')_{\times}.$\\

Each implies consistency inequality among the relevant correlations. This gives 16 consistency inequalities.

Eliminating the unmeasured correlations ($<AA'>, <BB'>$) gives the following inequalities

\begin{equation}
|<AB>\mp<A'B>| +|<AB'>\pm<A'B'>|\;\le\;2
\end{equation}

identified as Bell's inequalities. It is accepted that these inequalities are violated experimentally \cite{aspect}.
Thence an account of the EPRB set up by a canonical probability theory fails.
(No violation has been observed of the quantum mechanical (QM) predictions \cite{asher}.)

\section{The Quantum Mechanical State}

The culprit in the failure of an account of the EPRB set up by canonical probability theory appears to be the presence therein of
correlations among observables that do not have values simultaneously i.e. that can be revealed by the same experiment. (These are  A with A'  and B with B' in the EPRB set up. E.g. Eq.(\ref{simul}) assigns  instantaneous relation between values for A and A', within the same distribution.)  Such instantaneous relation are, we contend, disallowed by nature. Thence the distribution function
for values of, say, A must be different from that of A' (though both may pertain to the same quantum state) and their measurements must involve distinct experiments.\\

 The observed correlations violate the inequalities yet  abide by QM. This may seem unexpected in view of the analysis above since
the evaluation of the correlations within QM does involves the same quantum mechanical state, $\rho,$ for
all the correlations. This, seemingly, is equivalent to using  the same sample space and measure (i.e probability) in the
canonical probability formulation. Yet the later
led to consistency relations which are violated experimentally indicating thereby its inadequacy. However closer look at the
QM calculations show that {\it mathematical} attributes of the Hilbert space formalism {\it in effect} assigns distinct
( probability) distributions to non commuting observables. Thus in evaluating say the expectation of an arbitrary operator, $\hat{A}=\sum_{a}|a>a<a|$ for a state
specified by an arbitrary $\rho$ the effective $\rho$ is \cite{luo},\cite{r1} in effect (indicated by an arrow) diagonal,

\begin{equation}
\hat{\rho}\rightarrow \rho_{A}\equiv\sum_{a}|a><a|\rho|a><a|.
\end{equation}

Likewise, when evaluating $\hat{A'}$ , it is $\rho_{A'}\equiv\sum_{a'}|a'><a'|\rho|a'><a'|\neq \rho_{A}$, in general.
I.e. in effect we evaluate two non commuting observables with two different distribution functions though the QM state is the same.\\

These considerations were raised elsewhere \cite{r1} and led to viewing QM measurement as made of two stages. The first, termed
unrecorded measurement (URM), involves an unrecorded von Neumann  measurement. This is associated with, in general, an actual change
in the QM state. The second stage view the resultant state as amendable for a classical distribution account (for compatible observables) and
the uncovering of the final outcome is handled much like within  classical probability theory. Translated to the case at hand, the QM state
is viewed
as an information code  encoding  the possible distributions for the system \cite{newton}. The relevant distribution for a particular measurement
is  attained, by a measurement (this is illustrated below). Thus, whereas the classical state may be viewed as a set of values for attributes - the quantum state may
be viewed as an encoded set of possible distributions for these attributes. It "{\it is (the symbolic representation
of) the ensemble to which it belongs}"\cite{newton}. $\rho$ within a quantum measurement (i.e. Hilbert space formalism) is "projected" to classical like distribution for compatible observables. Once the distribution is determined the measurement's second stage
is purely classical: gaining the numerical value for, say, A within
the distribution $\rho_{A}$. Measurement selects  the  distribution \cite{lajos}, e.g.,

\begin{equation}
\hat{\rho}\rightarrow \;\rho_{A}\;(\textsf{via \;Unrecorded\;measurement})\;\rightarrow a \;(\textsf{via\; classical\; measurement}.)
\end{equation}

The interpretation has special significance when dealing with
two (or more) particles state that allows separate measurements for each: measurement of one particle discloses partial distribution, in general:

\begin{equation}
\rho\;\rightarrow\;\sum_{a,b,b'}|a>|b><b|<a|\rho|a>|b'><b'|<a|,
\end{equation}

where only the partial distribution of the first particle pertaining to the attribute $\hat{A}$ is revealed. In the special case wherein
$\rho$ relates to a (maximally) entangled state viewing the state as symbolic representation of an ensemble has conceptual advantage.
Thus, e.g., consider measuring
 the polarity along some axis, $\theta$, on particle 1 in a two photon maximally entangled state,

\begin{equation}
\rho =\frac{1}{2}\big[(|+>_1|+>_2+|->_1|->_2)(<-|_2<-|_1+<+|_2<+|_1)\big].
\end{equation}

The form of this state is independent of the direction of the polarizers. Thus rotating the polarizers to the $\theta$ direction
leaves the form of state invariant:

\begin{equation}\label{inv}
\rho\Rightarrow \rho_{\theta}=\frac{1}{2}\big[(|+,\theta>_1|+,\theta>_2+|-,\theta>_1|-,\theta>_2)(<+,\theta|_1<+,\theta|_2 + <-,\theta|_1<-,\theta|_2) \big].
\end{equation}

Undertaking a measurement along $\theta$ of one particle, gives:

\begin{equation}\label{dist}
\rho\Rightarrow \rho_{\theta}=\frac{1}{2}\big[|+,\theta>_1|+,\theta>_2<+,\theta|_1<+,\theta|_2 + |-,\theta>_1|-,\theta>_2<-,\theta|_2<-,\theta|_1\big] \big].
\end{equation}

I.e., in the case of a (maximally) entangled state, the one particle measurement (to uncover its distribution) revealed the complete distribution.

If the accessible particle is  measured yielding, say, +  the experimenter gained the knowledge that the remote particle is in the + state.\\

\section {Concluding Remarks}

A hidden variables account for the Einstein, Podolsky, Rosen and Bohm (EPRB) set up considered in the literature  \cite{aspect} is an
account in terms of classical (Kolmogorov's) probability theory. Such an account, necessarily, assigns values to all the EPRB two point
correlations both measured and unmeasured.\\

A mathematical attribute of standard quantum mechanics (QM)  (i.e. within its Hilbert space formulation) is an involvement of non commuting
observables. This may be viewed as reflecting  nature's disallowance for values of non-commuting observables to be simultaneously (within
the same experimental run) revealed. I.e. their having a defined instantaneous correlations within the same distribution is disallowed - they are incompatible.\\

QM accommodates the above mentioned natures' ruling,  classical probability theory in terms of hidden variables does not. This precludes an account via a canonical
probability theory of the EPRB set up: such an account implies, in principle, simultaneous values for some of these (un measured) incompatible  observables. Canonical probability formulation entails consistency requirements expressible as interrelation among correlations evaluated within the theory. They were identified as Bell's inequalities. We argued that their (experimental) violation   reflects classical probability theory  inadequacy in handling the incompatible observables.\\

Phase space formulation of QM aspire to provide it with classical like view. The formulation assigns the Wigner function the role of  distribution \cite{leonhardt}.( The role
of hidden variables is played by phase space coordinates.) A consistency condition for viewing the Wigner function as a classical like distribution is it be non negative. Thence the quantal demonstration of "violation" of positivity of the
Wigner function  is similar to  the quantal violation Bell's inequalities. In either case the direct classical like formalism fails
to uphold nature's disallowance of prescribed values for instantaneous correlations of incompatible observables. Both are
violated within QM that does uphold the disallowance.\\

An appropriate classical like  hidden variable account for the EPRB set up will, thus, require intricate information on ensemble of distributions   (which is incorporated
within the QM formalism). This is beyond the classical Kolmogorov probability theory. The issue of entanglement enters since for such states the full distribution is revealed via measurement of one of the constituents.\\

To summarize:\\

\noindent 1. A novel concise derivation of Bell's inequalities is presented. The derivation allow better isolation of the reason for their violation which support
viewing quantum measurement as two stage process and quantum states as encoding information on distributions in addition to their probabilistic attributes. \\

\noindent 2. The experimentally observed Bell's inequality violation does not relate in any obvious way to locality attribute of hidden variable account for Einstein, Podolsky, Rosen and Bohm (EPRB) set up. The violation does not rule out EPR contention that QM is an incomplete theory of real physical entities.\\

\noindent 3. Bell's inequalities are consistency conditions for classical probability theory. Their violation indicate inadequacy
of Kolmogorov's classical probability theory to account for the physics involved in EPRB set up. A proper account should require an extended theory, one that could deal with states encoding distributions of classical like distributions.\\

Acknowledgement: Numerous illuminating discussions  with Dr. Avi Levi, encouragement and  informative comments by
Professors P. Berman, K. Hess and L. Maccone  are gratefully acknowledged.\\

\end{document}